\documentclass{ws-ijmpe}
\usepackage[super,compress]{cite}
\usepackage{color}

\begin{document}

\markboth{Bao-Bao Jiao}{Nuclear binding energy predictions based on BP neural network}

\catchline{}{}{}{}{}
%%%%%%%%%%%%%%%%%%%%%%%%%%%%%%%%%%%%%%%%%%%%%%%%%%%%%%%%%%%%%%%%%%%%

\title{Nuclear binding energy predictions based on BP neural network}

\author{B. B. Jiao $^{1)2)}$}

\address{School of Nuclear Science and Engineering, East China University of Technology, Nanchang 330013, P. R. China$^{1)}$}
\address{Department of Physics, University of Shanghai for Science and Technology,
Shanghai 200093, P. R. China$^{2)}$\\
baobaojiao91@126.com}

\maketitle

\begin{history}
\received{Day Month Year}
\revised{Day Month Year}
%\accepted{Day Month Year}
%\comby{(xxxxxxxxxx)}
\end{history}

\begin{abstract}
Nuclear masses are of great importance in nuclear physics and astrophysics.
Descriptive experimental data on nuclear masses and the prediction of unknown masses based on residual proton-neutron interactions are a focus in nuclear physics.
The accuracy of the residual interaction determines the accuracy of the nuclear mass values,
so the study of residual interactions is essential.
Before we carry out this study, there are many papers using artificial neural networks in nuclear physics.
But no one uses BP neural network to study residual interactions.
In this paper, we obtained a description and prediction model for residual interactions based on BP neural network.
By combining experimental values with residual interactions model, we successfully calculate the nuclear masses of $A\geq100$.
Results demonstrate that the differences between our calculated values and experimental values (AME2003, AME2012 and AME2016) show that the root-mean-squared deviations (RMSDs) are small (comparing with AME2003, the odd-A nuclei RMSD and the even-A nuclei RMSD are 112 keV and 128 keV; comparing with AME2012,
the odd-A nuclei RMSD and the even-A nuclei RMSD are 103 keV and 121 keV; comparing with AME2016,
the RMSD of odd-A nuclei and even-A nuclei are 106 keV and 122 keV, respectively).
In addition, we obtained some predicted masses based on AME2003 and AME2012, the predicted values have good accuracy and compared well with experimental values (AME2012 and AME2016).
The results show that the study of residual interactions using the proposed BP neural network method is feasible and accurate.
This method is helpful for analyzing and extracting useful information from a large number of experimental values and then providing a reference for discovering physical laws and support for physical experiments.\\

\keywords{BP neural network; nuclear masses; residual proton-neutron interactions; binding energies.}
\end{abstract}

\ccode{PACS numbers:21.10.Dr; 21.45.Bc; 07.05.Mh}

%\tableofcontents

\section{Introduction}
The study of nuclear masses [1-22] is of great significance not only to nuclear physics, but also to astrophysics.
Early database (AME2003) [19] contained many experimental values of nuclear masses and many predictions of unknown masses,
the database published in 2017 (AME2016) [20] has some new nuclear masses than those published in 2012(AME2012)${[21]}$ and 2003.
In addition, the new measurement methods and the experimental instruments have made some nuclear masses more accurate.

In recent years, in addition to some global nuclear mass models [1-9], local mass relations have also attracted much attention.
For example: Audi-Wapstra systematics, Garvey-Kelson${[10]}$ local mass relations and the local mass relations of residual proton-neutron interactions [12-15, 22].
In Refs.[12-15, 22], the accuracy of residual interactions is closely related to the accuracy of nuclear masses, many methods and formulas have been developed to study the residual interaction in order to obtain accurate predicted values.
There are many papers using artificial neural networks in nuclear physics [23-33] and other subjects [34-37].
In the 1990s, people have used neural networks [23] to predict the mass of atomic nuclei.
Research in recent years, many improvements have been made based on the neural network approach to reduce the deviation of the calculated values or the predicted values [30-33].
The research in Ref. [30] shows that the accuracy of the Duflo-Zuker mass formula is significantly improved by using the Bayesian neural network approach, the RMSD relative to experiment is reduced from 503 keV to 286 keV (the accuracy is increased by about 40\% );
In Ref. [31], the prediction of nuclear mass obtained by Bayesian neural network method combined with Duflo-zuker model formula is compared with the newly experimental values in AME2016,
results show that the RMSD of prediction values and experimental values is about 400 KeV;
The Bayesian neural network approach in Ref. [32] can find the optimal value of the noise error automatically based on a distribution for the noise error in likelihood function, which reduces the deviation nuclear mass predictions;
Levenberg-Marquardt neural network approach [33] is used to study the nuclear masses in AME2012 database, results show that Levenberg-Marquardt neural network method is helpful to improve the accuracy of mass models, for a simple liquid drop formula: the RMSD between the predicted value and the 2353 experimental known masses decreased sharply from 2.455 MeV to 0.235 MeV, while for some other mass models, the accuracy is improved by about 30\%.
These facts show that neural network approach to improve the nuclear mass predictions of many models,
so we study the residual interaction of nuclei with $A\geq100$ based on BP neural network (using MATLAB2015b to build the model).
Based on the residual interaction proposed in Ref. [12] and the empirical formula to study the residual interaction, and inspired by the study of nuclear masses in Ref. [30], we use BP neural network to study the residual interaction.
The BP neural network is used to describe the residual interaction of nuclei with known masses and predict the residual interaction of nuclei with unknown masses, and then to describe and predict the binding energy (mass or mass excess) of nuclei.
Then obtained the RMSD of odd-A nuclei and even-A nuclei:
comparing with AME2003, RMSDs are 112 keV and 128 keV;
comparing with AME2012, RMSDs are 103 keV and 121 keV;
comparing with AME2016, RMSDs are 106 keV and 122 keV, respectively.
The predicted values based on databases (AME2003 and AME2012) are compared well with the experimental values.
A detailed understanding of the nuclear mass  for unknown mass that predicted value, is prerequisite to get more accurate nuclear mass map.

In this paper, we use BP neural network to study nuclear masses.
We describe and predict residual interactions based on the descriptive and predictive properties of BP neural network.
It is found that the RMSD of known masses nuclei has been reduced, and the accuracy of some nuclear mass predictions has also been improved.
In Sec. 2, the RMSD of known masses nuclei is calculated based on BP neural network and databases.
In Sec. 3, the predicted values of nuclear masses calculated by residual interactions and experimental values are compared with the experimental values. The comparison shows that our predicted values are close to the experimental values.
In Sec. 4, discusses this work.
In Sec. 5, summarizes this article.

\section{The residual interaction and BP neural network}

The residual interaction between nucleons is very important.
We assume that there is a nucleus with $Z$ protons and $N$ neutrons,
the relationship between binding energy $B(Z, N)$ and atomic mass $M(Z, N)$: $B(Z, N) = ZM_p +NM_n - M(Z, N)$,
where the $M_p$ and the $M_n$ are the mass of a free proton and a free neutron,
we then use a definition of the $B(Z,N)$ which yields a positive quantity and obtains a positive $\delta V_{1p-1n}$.
An experimental nuclear mass can be determined from a known atomic mass in AME databases.
The nuclear mass is equal to the atomic mass minus masses of electrons plus the electron binding energy.
The electron binding energy has little influence on the study of residual interactions. In addition, from the definition of residual interactions, we can see that the electron mass has no significance in residual interactions. Therefore, the electron binding energy and electron mass[38] are neglected in our studies. In this paper, we assume that the atomic mass is equal to the nuclear mass.

\subsection{the residual interaction}

The proton-neutron interactions between the last $i$ protons and $j$ neutrons is defined as[12]
\begin{eqnarray}
\label{eq:1}
\delta V_{ip-jn}(Z,N)=B(Z,N)+B(Z-i,N-j)-\nonumber\\[1mm]
B(Z,N-j)-B(Z-i,N).
\end{eqnarray}
It is easy to obtain the formula of $\delta V_{1p-1n}$,
\begin{eqnarray}
\label{eq:2}
\delta V_{1p-1n}(Z,N)=B(Z,N)+B(Z-1,N-1)-\nonumber\\[1mm]
B(Z,N-1)-B(Z-1,N).\nonumber\\[1mm]
\end{eqnarray}
we then get the mass equation as follows:
\begin{eqnarray}
\label{eq:3}
\ M(Z,N)=M(Z,N-1)+M(Z-1,N)- \nonumber\\[1mm]
M(Z-1,N-1)-\delta V_{1p-1n}.
\end{eqnarray}
\begin{eqnarray}
\label{eq:4}
\ M(Z,N)=M(Z+1,N)+M(Z,N+1)- \nonumber\\[1mm]
M(Z+1,N+1)-\delta V_{1p-1n}.
\end{eqnarray}
\begin{eqnarray}
\label{eq:5}
\ M(Z,N)=M(Z-1,N)+M(Z,N+1)- \nonumber\\[1mm]
M(Z-1,N+1)+\delta V_{1p-1n}.
\end{eqnarray}
\begin{eqnarray}
\label{eq:6}
\ M(Z,N)=M(Z+1,N)+M(Z,N-1)- \nonumber\\[1mm]
M(Z+1,N-1)+\delta V_{1p-1n}.
\end{eqnarray}
Eqs. (3)-(6) show that the accuracy of residual interactions determines the accuracy of nuclear masses.

\subsection{BP neural network}
Ref. [30] uses Bayesian neural networks to construct prediction models to study nuclear masses.
The Learning Process of Bayesian Neural Networks: (1) Based on the characteristics and experience of the problem, determine the types of network nodes and their value status; (2) determine the network structure by experience or training; (3) determine the conditional probability table for each node by experience or training,
for a comprehensive review, see Refs. [39-40].
The calculation of the conditional probability table in a Bayesian neural network method is an important part of using the proposed method to study nuclear masses.
Inspired by [23-33], our paper describes and predicts the residual interaction based on the BP neural network method, thus describing and predicting the nuclear masses.
BP (back-propagation) neural network [41-47] was proposed by Rumelhart and McClelland in 1986. Multi-layer feedforward neural network trained by error back propagation algorithm is the most representative and widely used neural network learning model.
According to the learning rule, BP neural network method can adjust the weighted value of the link chain to converge to the target through training.
The architecture of the BP neural network method is composed of several layers of interconnected neurons, usually including the input layer, output layer and several hidden layers, and each layer contains a number of neurons.
The network structure is shown in Fig. 1,
where the number of input values is 3, the number of hidden-layer units is 15, the number of output layers is 1, and the number of output values is 1.
\begin{figure}[tbp]
\begin{center}
\includegraphics[width=0.6\textwidth]{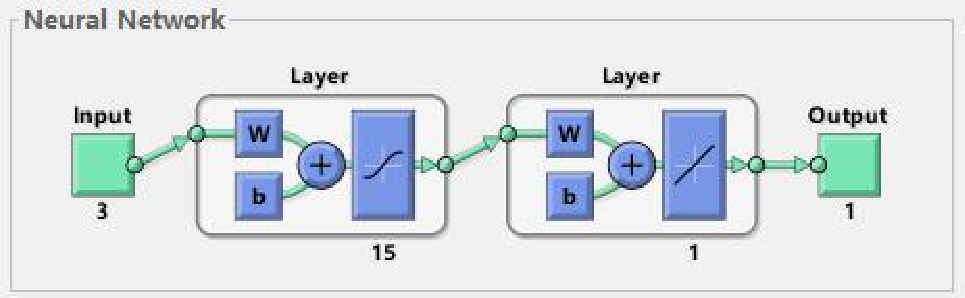}
\end{center}
\caption{(Color online) The BP neural network structure.}
\end{figure}

The learning process is as follows: (1) Select training data set; (2) Select a sample from the training data and input the information into the network; (3) Calculate the output of each layer node after neuron treatment; (4) Calculating the output error of the neural network and made a comparison with the setting error, if the requirement is met, the calculation stops, otherwise step 5 will continue; (5) From the output layer to the first hidden layer, the error will be reduced according to the principle, and the connection weights and thresholds of each neuron in the neural network will be adjusted; (6) Repeat (3) - (5) steps to the sample until the error of the training data meets the requirement.

In this paper, using the 'newff' function provided by the neural network toolbox of MATLAB 2015b to create the forward BP neural network, and then to study the proton-neutron residual interactions.
The main steps are as follows: with the proton number ($Z$), neutron number ($N$) and residual interactions as the input sample of the network, and set the output sample; obtained the BP network corresponding to the input and the output sample; network parameters are set as global minimum error $10^{-8}$, the number of intervals displayed 25, maximum training times 8000 and learning rate 0.03; then trained the network by using 'train' function; output the residual interaction.

We obtained the $ \delta V_{1p-1n}(Z,N)$ based on the database AME2003 (AME2012 and AME2016) and Eq. (2). The data sets Z, N, $ \delta V_{1p-1n}(Z,N)$ [calculated values of residual interaction for known mass] as the input sample(that is training data) for the network.
After training obtained the residual interaction $ \delta V_{1p-1n}(Z,N)$ [analog values of residual interaction of known mass] and $ \delta V_{1p-1n}(Z1,N1)$ [predicted value of residual interactions for unknown mass].
Using residual interactions and Eq. (3), combined with AME2003( AME2012 and AME2016), the values of the known nuclear masses were calculated and the values of unknown nuclear masses were predicted. Then, the RMSD between the calculated values of nuclear mass and the corresponding experimental values in AME2003(AME2012 and AME2016) was obtained.
The number of known-mass nuclei is different in AME2003, AME2012 and AME2016 databases.
Then, we obtain the three RMSDs in Table 1 by calculation.
Number0 corresponds to the number of nuclei in calculating the RMSDs.

\begin{table}
\centering
\caption{The RMSD of odd-A and even-A nuclei [obtained the RMSD based on BP neural network combined with databases
(AME2003,AME2012 and AME2016)], and shows the number of nuclei studied and the training times.}\label{tab1}
\vskip 2mm
\begin{tabular}{ccccc}
\hline  databases &  odd-even  & number0 &  training times  & RMSD (keV)  \\
\hline  AME2003   &odd         &    621           &1562               &$\simeq$112\\
                  &even        &    618           &3540               &$\simeq$128\\
\hline  AME2012   &odd         &    688           &1649               &$\simeq$103\\
                  &even        &    695           &3445               &$\simeq$121\\
\hline  AME2016   &odd         &    708           &2457               &$\simeq$106\\
                  &even        &    716           &2429               &$\simeq$122\\
\hline
\end{tabular}
\end{table}

From [12, 22] we can see that the residual interaction empirical formulas of odd-A nuclei are better than those of even-A nuclei.
Therefore, the RMSDs of odd-A nuclear masses are less than that of even-A nuclear masses, based on Eq. (3).
Table 1 also shows that the RMSD of odd-A nuclei is less than that of even-A nuclei when the nuclear mass is calculated by BP neural network.

In AME2003 database, for nuclei with mass number $A\geq100$,
Ref. [12] obtained the RMSD of odd-A nuclei (or even-A nuclei) between the calculated and experimental values of nuclei mass is 132 (or 168) keV.
In this paper, we study residual interactions based on the neural network function of tansig [$f(x)=2/(1+e^{-2x})-1$].
Because the neural network function comes from the neural network toolbox in MATLAB 2015b, so the main parameters of the function are known.
We obtained the nuclear mass based on [12] and the proposed BP neural network method. The results shows that the RMSD of odd-A nuclei is similar to that in [12], and the RMSD of even-A nuclei is smaller than that in [12], however, there are fewer parameters used in [12] than in our method.
In addition, we calculate nuclear masses $A\geq10$ and $A\geq42$, and the calculated values of the nuclear masses are close to the experimental values, which shows that using the BP neural network method to study residual interactions has advantages.
The comparison results show that our work is comparable to that of [12].

When we use the BP neural network method, the RMSDs obtained are compared with those in [12].
The advantage is that the nuclear masses studied in a larger region without increasing the deviation.
The disadvantages are that the method contains many parameters, and people cannot participate in the prediction process.
Table 2 shows that the RMSD of different mass regions.
RMSD1 corresponds to the root-mean-squared deviation for $A\geq10$, RMSD2 corresponds to the root-mean-squared deviation for $A\geq42$.
Number1 represents the number of nuclei in calculating the RMSD1,
number2 represents the number of nuclei in calculating the RMSD2.
In summary, the accuracy of describing residual proton-neutron interactions based on the proposed method is better than that using empirical formulas, which will be helpful in future astrophysical applications.

\begin{table}
\centering
\caption{The RMSD of odd-A and even-A nuclei for different mass regions [obtained the RMSD based on BP neural network combined with databases(AME2003, AME2012 and AME2016)], and shows the number of nuclei studied.}\label{tab2}
\vskip 2mm
\begin{tabular}{cccccc}
\hline  databases &  odd-even & number1  &   RMSD1 (keV) & number2  & RMSD2 (keV)  \\
\hline  AME2003   &odd        & 925           &224       & 834       &159\\
                  &even       & 915           &241       & 827       &167\\
\hline  AME2012   &odd        & 1018          &208       & 920       &132\\
                  &even       & 1024          &213       & 934       &157\\
\hline  AME2016   &odd        & 1051          &226       & 951      &142\\
                  &even       & 1063          &231       & 965      &167\\
\hline
\end{tabular}
\end{table}

\begin{figure}[tbp]
\begin{center}
\includegraphics[width=0.8\textwidth]{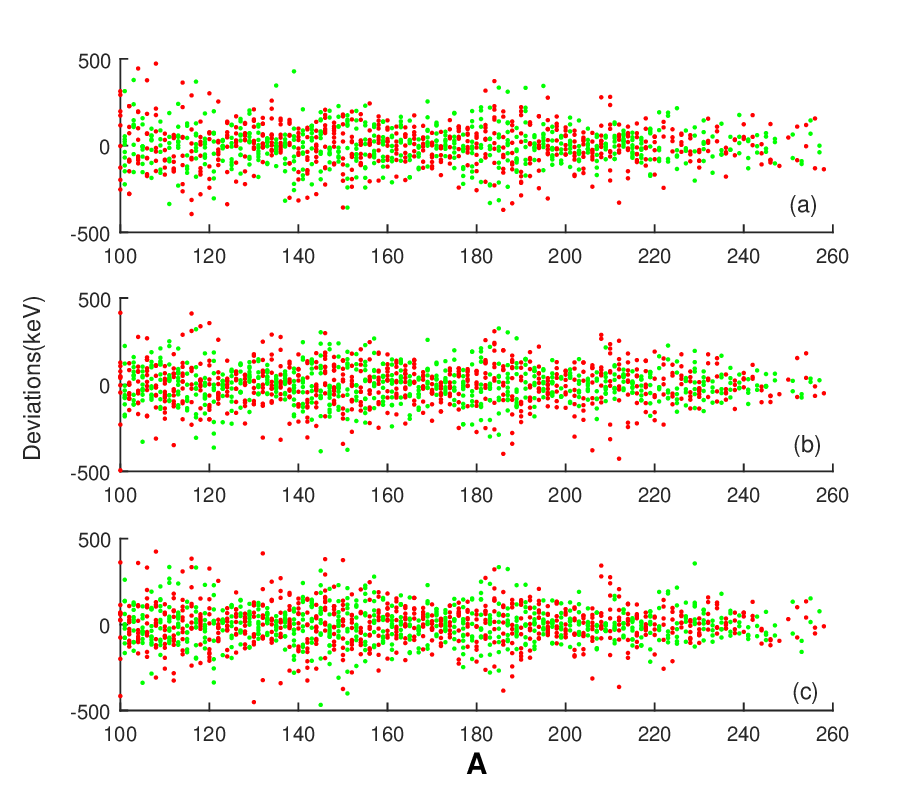}
\end{center}
\caption{(Color online) Deviations between our calculated values and the experimental values. In panels (a), (b), and (c), we present deviations separately for nuclei with even-A (red dots) and odd-A (green dots).}
\end{figure}

Fig. 2 shows the deviations between calculated and experimental values.
The dot in Fig. 2(a) represents the difference between the experimental value in the database AME2003 and the calculated value $M^{cal1}$ (we obtained the calculated value based on AME2003);
in Fig. 2(b) , dots show that the difference between the experimental value in the database AME2012 and the calculated value $M^{cal2}$(we obtained the calculated value based on AME2012);
in Fig. 2(c), dots are plotted by using the difference between the experimental value in the database AME2016 and the calculated value $M^{cal3}$(we obtained the calculated value based on AME2016).

We use the BP neural network method and Eqs. (4)-(6) with AME2012 database to study other kinds of $\delta V_{1p-1n}$ with mass number $A\geq100$.
The RMSD between the experimental values and calculated values is shown in Table 3.
Table 3 shows that the calculated values of nuclear masses are close to the experimental values.
These results demonstrate the feasibility and accuracy of using BP neural network to describe four kinds of residual interactions.
Therefore, we can predict the residual interaction of unknown masses nuclei on the basis of training model.

\begin{table}
\centering
\caption{The RMSD of odd-A and even-A nuclei(obtained the RMSD based on the  BP neural network combined
with AME2012 database and four formulas).}\label{tab3}
\vskip 2mm
\begin{tabular}{cccccc}
\hline  databases &  odd-even & Eq.(3)  &   Eq.(4) & Eq.(5)  & Eq.(6)  \\
\hline  AME2012   &odd        & 103 keV         &108 keV       & 112 keV       &116 keV\\
                  &even       & 121 keV          &133 keV       & 106 keV       &100 keV\\
\hline
\end{tabular}
\end{table}

In addition, we studied the $\delta V_{1p-2n}$ and $\delta V_{2p-1n}$ of nuclei (mass number $A\geq100$) based on AME2012 database and BP neural network method. From the calculation results, we can see that the calculated values of nuclear masses are close to the experimental values, and the RMSD is 126 keV and 149 keV respectively.
It is also known that the BP neural network can better describe and predict $\delta V_{1p-2n}$ and $\delta V_{2p-1n}$, and then get better prediction values of nuclear masses.
In this paper, we use Eqs. (3) and (4) to explain the effect of nuclear masses prediction. Of course, we can also use Eqs. (5) and (6) to predict nuclear masses.

\section{Prediction of Nuclear Masses}

In this section, the residual interaction of unknown mass nuclei are obtained based on the proposed method and AME2003 database.
Then, we predict unknown masses with Eqs. (3) and (4), and we calculate the average value
if the equations obtain the same nuclear mass.
In table 4 we present a set of selected data of mass excess of our predicted values and experimental values in the AME database (AME2012 and AME2016).

\begin{table}
  \centering
  \caption{The difference between the predicted values of nuclear mass (obtained by BP neural network and the AME2003
database) and experimental values in the AME2012 and AME2016 database.}\label{tab4}
  \vskip 2mm
\begin{tabular}{cccccc}
\hline											
Nucleus	&	ME2012	&	ME2016	&	ME$^{pred1}$	&	dev1	&	dev2	\\
	&	/keV	&	/keV	&	/keV	&	/keV	&	/keV	\\
\hline
$^{99}Y$	&	-70656	&	-70650	&	-70249	&	-407	&	-401	\\
$^{99}Zr$	&	-77624	&	-77621	&	-77522	&	-102	&	-99	\\
$^{99}Nb$	&	-82332	&	-82335	&	-82563	&	231	&	228	\\
$^{99}Mo$	&	-85969	&	-85970	&	-85838	&	-131	&	-132	\\
$^{99}Tc$	&	-87327	&	-87328	&	-87299	&	-28	&	-29	\\
$^{99}Ru$	&	-87622	&	-87625	&	-87453	&	-169	&	-172	\\
$^{99}Rh$	&	-85578	&	-85581	&	-85457	&	-121	&	-124	\\
$^{99}Pd$	&	-82181	&	-82183	&	-82187	&	6	&	4	\\
$^{99}Ag$	&	-76712	&	-76712	&	-76588	&	-124	&	-124	\\
$^{103}Y$	&	-58458	&	-58458	&	-58912	&	454	&	454	\\
$^{119}Pd$	&	-71408	&	-71408	&	-71091	&	-317	&	-317	\\
$^{129}Cd$	&	-63510$^{The.}$	&	-63058	&	-62943	&	Null	&	-115	\\
$^{147}Ba$	&	-60264	&	-60264	&	-60617	&	353	&	353	\\
$^{149}Ba$	&	-53020$^{The.}$	&	-53120	&	-53211	&	Null	&	91	\\
$^{155}Nd$	&	-62284	&	-62284	&	-62444	&	160	&	160	\\
$^{161}Ta$	&	-38701	&	-38779	&	-38885	&	184	&	106	\\
$^{106}Sb$	&	-66473	&	-66473	&	-66541	&	68	&	68	\\
$^{108}Sb$	&	-72445	&	-72445	&	-72367	&	-78	&	-78	\\
$^{110}Sb$	&	-77450	&	-77450	&	-77693	&	243	&	243	\\
$^{110}I$	&	-60460	&	-60460	&	-60732	&	272	&	272	\\
$^{112}I$	&	-67063	&	-67063	&	-66947	&	-116	&	-116	\\
$^{114}Ru$	&	-70222	&	-70222	&	-70688	&	466	&	466	\\
$^{114}Cs$	&	-54680	&	-54680	&	-55204	&	524	&	524	\\
$^{142}Tb$	&	-56560	&	-56560	&	-56688	&	128	&	128	\\
$^{160}Eu$	&	-63480	&	-63480	&	-63613	&	133	&	133	\\
$^{164}Re$	&	-27520	&	-27470	&	-27388	&	-132	&	-82	\\
$^{166}Re$	&	-31890	&	-31890	&	-31737	&	-153	&	-153	\\
$^{168}Ir$	&	-18720	&	-18670	&	-18525	&	-195	&	-145	\\
$^{172}Ir$	&	-27380	&	-27380	&	-27269	&	-111	&	-111	\\
$^{172}Au$	&	-9370	&	-9320	&	-8979	&	-391	&	-341	\\
$^{176}Au$	&	-18400	&	-18520	&	-18277	&	-123	&	-243	\\
$^{180}Tl$	&	-9260	&	-9390	&	-9309	&	49	&	-81	\\
$^{228}Fr$	&	33369	&	33384	&	33048	&	321	&	336	\\
$^{228}Np$	&	33600	&	33600	&	33879	&	-279	&	-279	\\
\hline
\end{tabular}
\end{table}

There are more than 214 new nuclei appearing in the AME2012 database compared to the AME2003 database, with approximately 168 new nuclei $A\geq99$. In addition, calculating the new nuclear masses requires three known nuclei, so the predicted number of nuclei will be smaller than 168. We obtained 58 predicted mass values of nuclei based on the AME2003 database, and 34 predicted values were given experimental values in AME2012 and AME2016 database.
In Table 4, we give 34 predicted values in comparison with the experimental values.
Where, dev1 represents the difference between the experimental values (ME2012) of the mass excess in the AME2012 database and our predicted values (ME$^{pred1}$);
we use dev2 to represent the difference between the experimental values (ME2016) of the mass excess in the AME2016 database and our predicted values (ME$^{pred1}$);
These experimental values given in AME2012 and AME2016 are rounded.
From the above table, we can see that the predicted value is close to the experimental value, and some nuclear mass deviations are only tens of keV or even keV.
In addition, two nuclei ($^{129}Cd$ and$^{149}Ba$) have only predicted values in the AME2012 database without experimental values, it is found that our predicted value is closer to the experimental value (AME2016) than the predicted value in AME2012.
Although the deviation of some nuclear masses larger than is desired, it has little effect on the overall description and prediction of nuclear masses.
The residual interaction based on BP neural network method is inaccurate, which leads to a deviation between the calculated value and the experimental value.
Therefore, the smaller the residual interaction deviation is, the more accurate the nuclear mass will be.

In addition, we used the BP neural network method and the AME2012 database to predict unknown masses.
We calculate the average value if Eqs. (3) and (4) obtain the same nuclear mass.
Table 5 shows some mass excess of the experimental values (AME2016) compared with the predicted values,
where dev3 represents the measurement deviation of nuclear mass in the AME2016 database,
dev4 represents the difference between the experimental values(ME2016) in AME2016 database and our predicted values (ME$^{pred2}$).
The results show that the predicted values are in good agreement with the experimental values (AME2016).

\begin{table}
  \centering
  \caption{The difference between the predicted values of nuclear mass (obtained by BP neural network and the AME2012
database) and experimental values in the AME2016 database.}\label{tab5}
  \vskip 2mm
\begin{tabular}{ccccc}
\hline									
Nucleus	&	ME2016(keV)	&	dev3(keV)	&	ME$^{pred2}$(keV)	&	dev4(keV)	\\
\hline									
$^{99}Y$	&	-70650	&	7	&	-70557	&	-93	\\
$^{99}Zr$	&	-77621	&	11	&	-77349	&	-272	\\
$^{99}Nb$	&	-82335	&	12	&	-82515	&	180	\\
$^{99}Mo$	&	-85970	&	0.23	&	-85877	&	-93	\\
$^{99}Tc$	&	-87328	&	0.9	&	-87274	&	-54	\\
$^{99}Ru$	&	-87625	&	0.3	&	-87379	&	-246	\\
$^{99}Rh$	&	-85581	&	7	&	-85373	&	-208	\\
$^{99}Pd$	&	-82183	&	5	&	-82041	&	-142	\\
$^{99}Ag$	&	-76712	&	6	&	-76532	&	-180	\\
$^{129}Cd$	&	-63058	&	17	&	-63369	&	311	\\
$^{141}I$	&	-59927	&	16	&	-59824	&	-103	\\
$^{149}Ba$	&	-53120	&	440	&	-52937	&	-183	\\
$^{190}Tl$	&	-24372	&	8	&	-24432	&	60	\\
$^{194}Bi$	&	-16029	&	6	&	-15985	&	-44	\\
$^{198}At$	&	-6715	&	6	&	-6653	&	-62	\\
$^{202}Fr$	&	3096	&	7	&	3164	&	-68	\\
\hline
\end{tabular}
\end{table}

\section{Results and discussion}
In this paper we study systematics of the residual interaction.
Sec. 2 and Sec. 3 obtain the description values and prediction values of the nuclear mass.
These results show that the method based on BP neural network to describe and predict the residual interaction is feasible.
The RMSD of the known mass nuclei is reduced;
at the same time, the predicted values based on the residual interaction and the database (AME2003 and AME2012) are also close to the experimental values.
The advantage of BP neural network method is that the nuclear masses studied in a larger region without increasing the deviation.
The disadvantage is that we canot participate in the operation of the neural network.

Our calculated values based on BP neural network are in good agreement with the experimental values.
In addition, the residual interaction of odd-A nuclei is statistically good.
Therefore, the RMSD of odd-A nuclei is less than that of even-A nuclei when describing the known masses.
As shown in Table 4 (Table 5), the predicted values based on AME2003 (AME2012) are close to the experimental values in AME2012 and AME2016 (AME2016),
and some of the deviations between predicted values and experimental values are only tens of keV.
In summary, BP neural network method can describe the AME database (AME2003, AME2012, AME2016) and predict unknown masses.

Both Ref. [30] and our paper use neural networks to study nuclear masses.
Ref. [30] uses Bayesian neural network to study nuclear masses based on the global mass relations (Duflo-Zuker mass formula). Results show that the root-mean-squared deviation relative to experiment is reduced from 503keV to 286keV. The Bayesian neural network approach is highly successful in refining the predictions of existing mass models.
In this paper, we use BP neural network method to study nuclear masses based on local mass relations (residual proton-neutron interactions mass formula). It is found that the RMSD of known masses is less than that obtained by empirical formula, and the accuracy of some unknown masses has been improved.
In addition, the region of masses that can be studied is larger.
Bayesian network [48-50] is a graphical model, which describes the relationship between related random variables or events.
This algorithm is also a model for reasoning.
It uses a directed acyclic graph to represent the causal relationship between random variables or events, and the relationship between these nodes is quantified by a conditional probability table.
Back Propagation (BP) neural network is one kind of multilayer feed-forward Neural Network. It is usually trained by back propagation of errors. The results show that the deviation of calculation values and prediction value is reduced by using the BP neural network and Bayesian neural network.

\section{Conclusions}
The main work of this paper is to study residual interactions ($\delta V_{1p-1n}$).
Previously, many researchers were using artificial networks in nuclear physics.
In our paper, we proposed use BP neural network method to build description and prediction models based residual interactions (odd-A nuclei and even-A nuclei, respectively).
We then describe and predict nuclear masses using the residual interaction combined with experimental values.
Although the deviation of some nuclear masses larger than is desired, it does not affect the overall description and prediction effect.

When using BP neural network method for data fitting, as long as we select an appropriate network structure, we can obtain a better fitting curve and meet the requirements of different users.
The BP neural network method in this paper is suitable for medium and long term prediction.
It has the advantages of fast calculation speed, high precision and strong nonlinear fitting ability,
which the nuclear masses studied in a larger region without increasing the deviation in describing and predicting nuclear masses.
Because it cannot participate in the prediction process, the algorithm is incomplete, these factors can lead to deviation.
Our predicted values of some unknown masses will be useful in future studies.
The more precise residual interactions, the more accurate nuclear masses will be obtained.

\section*{Acknowledgements}
The author would like to thank L. Y. Jia for reading and commenting of this paper.
This work is supported by the Doctoral Scientific Research Foundation of East China University of Technology (DHBK2019151).


\begin{thebibliography}{20}
\bibitem{lab1}
C F Von Weizs$\ddot{\mathrm{a}}$cker {\it Z. Phys} {\bf 96} 431 (1935)
\bibitem{lab2}
P M$\ddot{\mathrm{o}}$ller, J R Nix, W D  Myers et al {\it At. Data Nucl. Data Tables} {\bf 59} 185 (1995)
\bibitem{lab3}
L Geng, H Toki and J Meng {\it Prog. Theor. Phys.} {\bf 113} 785 (2005)
\bibitem{lab4}
S Goriely, F Tondeur and J M Pearson {\it At. Data Nucl. Data Tables} {\bf 77} 311 (2001)
\bibitem{lab5}
S Goriely, N Chamel and J M Pearson {\it Phys. Rev. Lett.} {\bf 102} 152503 (2009)
\bibitem{lab6}
P M$\ddot{\mathrm{o}}$ller, W D Myers, H Sagawa et al {\it Phys. Rev. Lett.} {\bf 108} 052501 (2012)
\bibitem{lab7}
C Qi  {\it J. Phys. G: Nucl. Par.}  {\bf 42} 045104 (2015)
\bibitem{lab8}
N Wang, Z Liang, M Liu et al {\it Phys. Rev. C} {\bf 82} 044304 (2010)
\bibitem{lab9}
J Duflo and A P Zuker {\it Phys. Rev. C} {\bf 52} R23 (1995)
\bibitem{lab10}
G T Garvey and I Kelson {\it Phys. Rev. Lett.} {\bf 16} 197 (1966)
\bibitem{lab11}
M Thoennessen {\it Int. J. Mod. Phys. E} {\bf27} 1830002 (2018)
\bibitem{lab12}
G J Fu , Y Lei, H Jiang et al {\it Phys. Rev. C} {\bf 84} 034311 (2011)
\bibitem{lab13}
H Jiang, G J Fu, B Sun et al {\it Phys. Rev. C} {\bf 85} 054303 (2012)
\bibitem{lab14}
G J Fu, H Jiang, Y M Zhao et al {\it Phys. Rev. C} {\bf 82} 034304 (2012)
\bibitem{lab15}
H Jiang, G J Fu, Y M Zhao et al {\it Phys. Rev. C} {\bf 82} 054317 (2010)
\bibitem{lab16}
D Lunney, J M Pearson and C Thibault {\it Rev. Mod. Phys.} {\bf 75} 1021 (2003)
\bibitem{lab17}
B B Jiao {\it Mod. Phys. Lett. A} {\bf 32} 1850156 (2018)
\bibitem{lab18}
Z Wu, S A Changizi and C Qi {\it Phys. Rev. C}  {\bf 93} 034334 (2016)
\bibitem{lab19}
G Audi, A H Wapstra and C Thibault {\it Nucl. Phys. A} {\bf 729} 337 (2003)
\bibitem{lab20}
G Audi,  F G Kondev, M Wang et al {\it Chin. Phys. C} {\bf 41} 030001 (2017)
\bibitem{lab21}
M Wang, G Audi, A H Wapstra et al {\it Chin. Phys. C} {\bf 36} 1603 (2012)
\bibitem{22}
B B Jiao {\it Sci Sin-Phys Mech Astron} {\bf 48} 052001 (in Chinese) (2018)
\bibitem{23}
Gazula S, Clark J W and Bohr H {\it Nucl. Phys. A}  {\bf540} 1 (1992)
\bibitem{24}
Gernoth K A, Clark J W, Prater J S et al {\it Phys. Lett. B}  {\bf300} 1 (1993)
\bibitem{25}
T Bayram, S Akkoyun and S O Kara {\it Ann. Nucl. Energy} {\bf63} 172 (2014)
\bibitem{26}
S Athanassopoulos, E Mavrommatis, K A Gernoth, et al  {\it Nucl. Phys. A}  {\bf743} 222 (2004)
\bibitem{27}
J W Clark and H Li {\it Int. J. Mod. Phys. B} {\bf20} 5015 (2006)
\bibitem{28}
N J Costiris, E Mavrommatis, K A Gernoth, et al {\it Phys. Rev. C} {\bf80} 044332 (2009)
\bibitem{29}
S Akkoyun, T Bayram and T Turker {\it Radiat. Phys. Chem.} {\bf96} 186 (2014)
\bibitem{30}
R Utama and J Piekarewicz {\it Phys. Rev. C} {\bf 96} 044308 (2017)
\bibitem{31}
R Utama and J Piekarewicz {\it Phys. Rev. C} {\bf 97} 014306 (2018)
\bibitem{32}
Z M Niu and H Z Liang {\it Physics Letters B} {\bf 778} 48 (2018)
\bibitem{33}
H F Zhang, L H Wang, J P Yin, et al {\it J. Phys. G: Nucl. Par.}{\bf44} 045110 (2017)
\bibitem{34}
J He, X Tang, P Gong, et al {\it Ann. Nucl. Energy} {\bf 112} 1 (2018)
\bibitem{35}
D Ma, T Zhou, J Chen, et al {\it Nucl. Eng. Des.} {\bf 320} 400 (2017)
\bibitem{36}
D Z Wei, F J Chen and X X Zhen {\it Acta Phys. Sin.} {\bf 64} 110503 (in Chinese) (2015)
\bibitem{37}
K X Peng, J B Yang, X G Tuo, et al {\it Mod. Phys. Lett. B } {\bf 30} 87 (2016)
\bibitem{38}
J Tang, Z M Niu and J Y Guo {\it Chin. Phys. C} {\bf40} 074102 (2016)
\bibitem{39}
C Bishop, Neural Networks for Pattern Recognition (Oxford University Press, Birmingham, UK, 1995).
\bibitem{40}
S Haykin, Neural Networks: A Comprehensive Foundation (Prentice Hall, Upper Saddle River, NJ, 1999).
\bibitem{41}
I Kaastra and M Boyd {\it Neurocomputing} {\bf10} 215 (1996)
\bibitem{42}
A Parlak, Y Islamoglu, H Yasar, et al {\it Appl. therm. eng. } {\bf26} 824 (2006)
\bibitem{43}
G Jing, W Du and Y Guo {\it Desalination } {\bf 291} 78 (2012)
\bibitem{44}
 B H M Sadeghi {\it J. Mater. Process. Technol.} {\bf 103} 411 (2000)
\bibitem{45}
L L Zhao, Y X Yang, Y M Zheng, et al {\it Journal of East China Institute of Technology} {\bf36} 79 (2013)
\bibitem{46}
H Z Xu, S Wu and N Liu {\it Journal of East China Institute of Technology} {\bf3} 293 (2008)
\bibitem{47}
X X Cheng, X Y Cheng and C J Liu {\it Journal of East China Institute of Technology} {\bf39} 172 (2016)
\bibitem{48}
R Utama, J Piekarewicz and H B Prosper {\it Phys. Rev. C} {\bf 93} 014311 (2016)
\bibitem{49}
L Alvarez-Ruso, K M Graczyk and E Saul-Sala {\it Phys. Rev. C} {\bf 99} 025204 (2019)
\bibitem{50}
S Arangio and F Bontempi {\it Struct. Infrastruct. Eng.} {\bf 11} 575 (2015)
\end{thebibliography}
\end{document}